\documentclass[amsmath]{aastex702}

\begin{document}

   \title{Ultraviolet accretion continua reveal a transition in T Tauri magnetic topology}

%
%
%

\author[,sname='G\'oomez de Castro']{Ana I. G\'omez de Castro }
\altaffiliation{AEGORA Research Group -- Joint Center for UV Astronomy, Universidad Complutense de Madrid}
\affiliation{Departamento de Fisica de la Tierra y Astrof\'{\i}sica, Fac. de CC. Matem\'aticas, Universidad Complutense de Madrid}
\email[show]{aig@ucm.es}

 
  \begin{abstract}
 
Magnetic fields regulate accretion in T Tauri stars (TTSs) and evolve as radiative cores emerge during pre-main-sequence contraction. Stars above $\sim$0.35--0.5 M$_{\odot}$ transition from fully convective to partially convective interiors, a change associated with increasingly complex magnetic topologies and evolving dynamo behaviour. Because accretion flows are magnetically channelled, this evolution is expected to modify ultraviolet (UV) accretion diagnostics, although observational evidence has remained limited.

We identify a transition in the UV continuum properties of TTSs using archival \textit{International Ultraviolet Explorer} observations, complemented by recent spectra from the Hubble Space Telescope ULLYSES survey. An empirical UV photospheric sequence constructed from G--M main-sequence stars shows that late K--M TTSs broadly follow standard photosphere-plus-shock model predictions, whereas G--K5 stars display systematically lower $F270/F303$ continuum flux ratios (flux at 270~nm relative to 303~nm).

Statistical analysis shows that transition models describe the observations significantly better than a single continuous trend, locating the transition near spectral type K5--K6, with its precise location limited by spectral-classification uncertainties, and consistent with  the fully convective boundary. Comparison with photosphere-plus-shock models further shows that, while the late-type population is broadly encompassed by the parameter space sampled by the adopted models, the warmer stars extend systematically beyond it in this UV diagnostic.

We interpret this transition as evidence  for a change in magnetospheric accretion associated with  the evolution of magnetic topology across the radiative-core boundary, resulting in a corresponding change in the emergent UV continuum.
  
    \end{abstract}
   
\keywords{T Tauri stars (1681) --- Pre-main sequence stars (1290) --- Stellar accretion (1578) --- Stellar magnetic fields (1610) --- Ultraviolet astronomy (1736)}

\section{Introduction}

Magnetic fields regulate the interaction between T Tauri stars (TTSs) and their circumstellar disks through magnetospheric accretion, in which gas from the inner disk is channelled along magnetic field lines onto the stellar surface \citep{1991ApJ...370L..39K,1994ApJ...426..669H,2007prpl.conf..479B,2016ARA&A..54..135H}. The resulting accretion shocks generate strong ultraviolet (UV) excess emission, making the UV continuum one of the most sensitive diagnostics of the accretion flow \citep{2007ApJ...664..975J,2009ARA&A..47..333D,2012ApJ...755...97G,1996MNRAS.283...55G,1999MNRAS.304L..41G,2000ApJ...544..927G,2025ApJ...992..134P}. During pre-main-sequence evolution, stars more massive than $\sim0.35$--$0.5,M_\odot$ develop radiative cores. This transition is accompanied by profound changes in magnetic topology, from predominantly strong dipolar fields to weaker, more complex magnetic configurations inferred from spectropolarimetric observations and dynamo models \citep{2024MNRAS.530..264D,2026A&A...707A..45G,2016MNRAS.457.3836G}. Such changes are expected to modify the geometry of magnetospheric accretion and, consequently, the properties of the resulting accretion shocks \citep{2007MNRAS.374..436L}.

Standard magnetospheric accretion models successfully reproduce many UV and optical properties of classical TTSs \citep{2013ApJ...767..112I,2018MNRAS.475.4367D}. However, it remains unknown whether the observed UV continuum properties of accreting TTSs exhibit systematic changes across the transition from fully convective to partially convective stars. Addressing this observational question requires empirical photospheric references because current stellar-atmosphere models do not accurately reproduce the UV spectra of cool stars.

In this Letter we investigate whether the ultraviolet continuum properties of accreting TTSs exhibit a systematic transition with spectral type using homogeneous ultraviolet observations of 68 accreting TTSs obtained with the \textit{International Ultraviolet Explorer} (IUE) and the HST/ULLYSES survey \citep{2025ApJ...985..109R,1997iueu.book.....G}.

We define an extinction-insensitive UV continuum colour diagnostic based on the $F_{270}/F_{303}$ flux ratio and compare it with empirical photospheric colours and a grid of accretion-shock models. Rather than modelling the global UV--optical spectra of individual stars, we use the model grid as a physical reference for interpreting the behaviour of this observational diagnostic across spectral type.  We find a statistically significant transition in the ultraviolet continuum properties of accreting TTSs near spectral type K5--K6. Comparison with empirical photospheric colours and the parameter space sampled by the adopted accretion-shock models shows that late K--M stars are broadly encompassed by the models, whereas many warmer G--K stars extend beyond this range towards systematically weaker UV excesses. We interpret this observational transition in the context of the well-established evolution of stellar magnetic topology.

\section{Accretion shocks and UV colours}

To investigate the ultraviolet continuum produced by magnetospheric accretion, we define two colour indices that characterise the continuum shape while minimising contamination from strong emission lines. The observed UV continuum is the sum of the stellar photosphere, thermal emission from the heated photosphere beneath the accretion shock, and Balmer continuum emission from the photoionized pre-shock gas \citep{2013ApJ...767..112I,2018MNRAS.475.4367D}. Continuum windows centred at 270, 290 and 303\,nm were selected to sample these components while avoiding the strongest emission features in T Tauri spectra. Together with the $V$ band, they define the two colour indices used throughout this work. The purely UV colour $\log(F270/F303)$ measures the slope of the Balmer continuum and provides an extinction-insensitive diagnostic of the relative contribution of the accretion continuum. The UV--optical colour $\log(F290/F_V)$ combines the near-UV continuum with the stellar photosphere and is primarily sensitive to the strength of the accretion continuum through its dependence on both shock temperature and fractional surface coverage (Fig.~1).

\begin{figure*}[ht!]
\centering
\begin{tabular}{cc}
\includegraphics[width=0.50\textwidth]{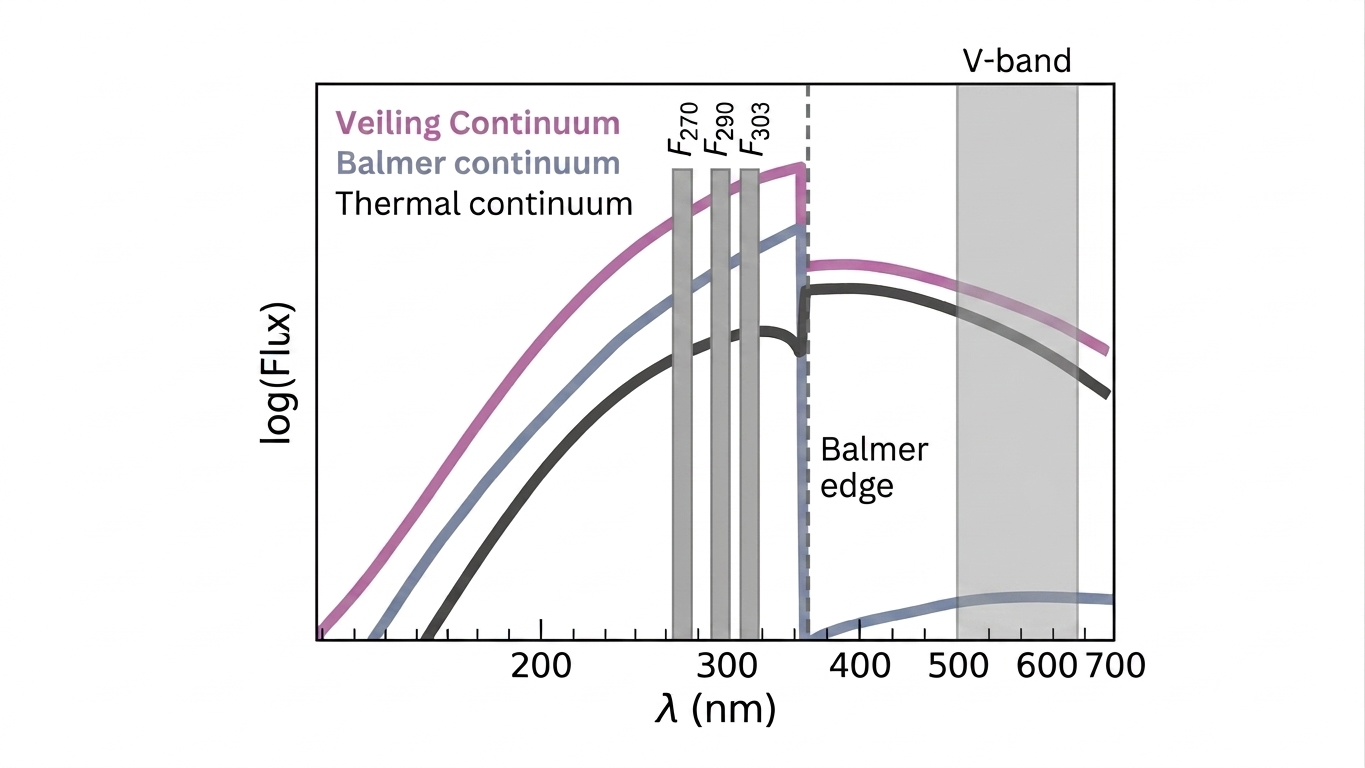} &
\includegraphics[width=0.50\textwidth]{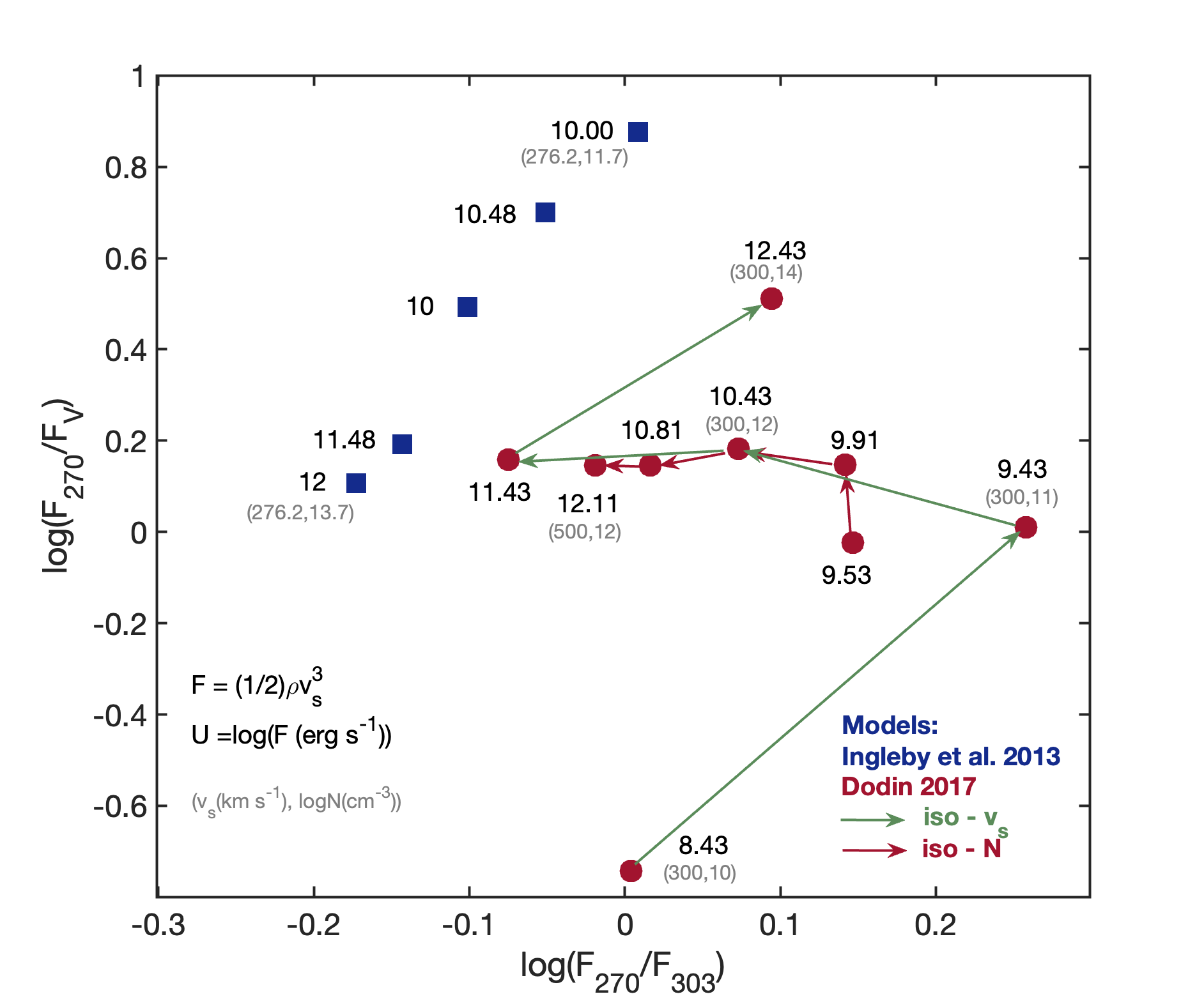}
\end{tabular}
\caption{UV continuum diagnostics and accretion-shock models.
\textbf{Left,} Schematic illustration of the continuum components contributing to the UV excess in accreting TTSs.
\textbf{Right,} Predicted locations of accretion-shock models in the $\log(F270/F303)$--$\log(F290/F_V)$ plane. Blue squares show the models of \citet{2013ApJ...767..112I}, labelled by the accretion energy flux ($\log \mathcal{F}$), with representative combinations of infall velocity ($v_{\rm s}$) and pre-shock density ($N$) indicated for selected models. Red circles show the models of \citet{2018MNRAS.475.4367D}. Green arrows denote increasing infall velocity at fixed density, and red arrows increasing density at fixed velocity. The diversity of model trajectories illustrates how the UV continuum colours respond to variations in the physical properties of magnetospheric accretion flows, in particular to changes in pre-shock density and infall velocity.
}
\label{fig:models}
\end{figure*}

To provide a physical reference for the observed UV continuum colours, we constructed a grid of photosphere-plus-shock models spanning spectral types G0--M4. Accretion-shock continua from fourteen published shock models were added to the corresponding empirical photospheric spectra. The strength of the accretion continuum was parameterised by the optical veiling,

\begin{equation}
R=\frac{F_{\rm veil}(V)}{F_\star(V)},
\end{equation}

where $F_{\rm veil}(V)$ and $F_\star(V)$ are the veiling and photospheric fluxes in the $V$ band. Models were computed for veiling factors $R=10^{-3}$--0.5, spanning the range from weakly to strongly accreting T Tauri stars. The ensemble of models defines the photosphere-plus-shock envelope used for comparison with the observations in Fig.~3.  This envelope represents the range of UV colours spanned by the adopted combinations of shock and photospheric parameters and is not intended to delimit the full parameter space accessible to magnetospheric accretion-shock models.

The UV photospheric sequence for G--M main-sequence stars was constructed from archival HST and IUE spectra rather than synthetic stellar atmospheres because current atmosphere models systematically underestimate the near-UV continuum of cool stars. The resulting photospheric sequence defines the observational baseline against which the T Tauri colours are compared, while the photosphere-plus-shock models provide a physical framework for interpreting the observed behaviour.

\section{Data}

Our TTS sample combines archival ultraviolet spectra from the \textit{International Ultraviolet Explorer} (IUE) with recent observations from the HST/ULLYSES survey. All available long-wavelength IUE prime and redundant spectra of classical TTSs were inspected, and observations with insufficient signal-to-noise for reliable continuum measurements were discarded. These observations were complemented with HST/STIS spectra from the ULLYSES survey for which reliable contemporaneous V-band fluxes could be derived from the optical spectra, providing homogeneous coverage of the ultraviolet continuum over 200--320 nm. The final sample comprises 68 accreting TTSs.

Ultraviolet fluxes were measured in three 5-nm-wide continuum windows centred at 270, 290 and 303\,nm. Optical fluxes were obtained from the IUE Fine Error Sensor (FES) or from synthetic $V$-band photometry derived from the HST spectra.

Several stars were observed repeatedly over timescales ranging from days to decades. For stars with multiple observations, colours were derived from regressions between correlated fluxes whenever statistically justified; otherwise, individual colour measurements were averaged. Although the sources exhibit substantial UV variability, the fluxes measured at 270 and 303\,nm remain tightly correlated, making the $\log(F270/F303)$ colour largely insensitive to variability. The close agreement between the colours independently derived from the IUE and HST observations of BP Tau, obtained approximately 40 years apart, illustrates the long-term stability of this diagnostic. The properties of the TTS sample are listed in Table~1.

Interstellar extinction values were adopted from the ULLYSES database or, when unavailable, from the IUE TTS atlas. Extinction corrections were computed using a standard Galactic extinction law with $R_V=3.1$. Because the $F270/F303$ colour compares nearby wavelengths, reddening introduces only minor corrections and does not affect the conclusions of this work.

\subsection{Main-sequence comparison sample}

To establish the photospheric reference sequence, we analysed archival ultraviolet spectra of non-accreting G--M dwarfs obtained with IUE and HST, including observations from the Next Generation Spectral Library and CALSPEC. Ultraviolet fluxes were measured using the same continuum windows adopted for the TTS sample to ensure a homogeneous comparison \citep{2012RSPTA.370.2765A,2022ApJ...933..235P}.

The photospheric reference sequence was constructed directly from the observations rather than from synthetic stellar atmospheres because current atmosphere models systematically underestimate the near-ultraviolet continuum of cool stars. The main-sequence sample was grouped into spectral-type bins, and the median $\log(F270/F303)$ values were interpolated to define the reference sequence shown in Fig.~1. The properties of the comparison stars are listed in Supplementary Table~2.

\section{TTSs in the UV-colour map}

The distribution of TTSs in the UV colour plane is shown in Fig.~2. Late K--M stars define a well-populated sequence that closely follows the loci {\bf spanned by the adopted photosphere-plus-shock model grid}. In contrast, warm G--K5 stars occupy a distinct locus, with systematically smaller $\log(F270/F303)$ values for comparable $\log(F290/F_V)$ colours. This behaviour is independently recovered in both the IUE and HST/ULLYSES samples, indicating that it is not produced by instrumental effects or sample selection.
\begin{figure}
\centering
\includegraphics[width=0.50\textwidth]{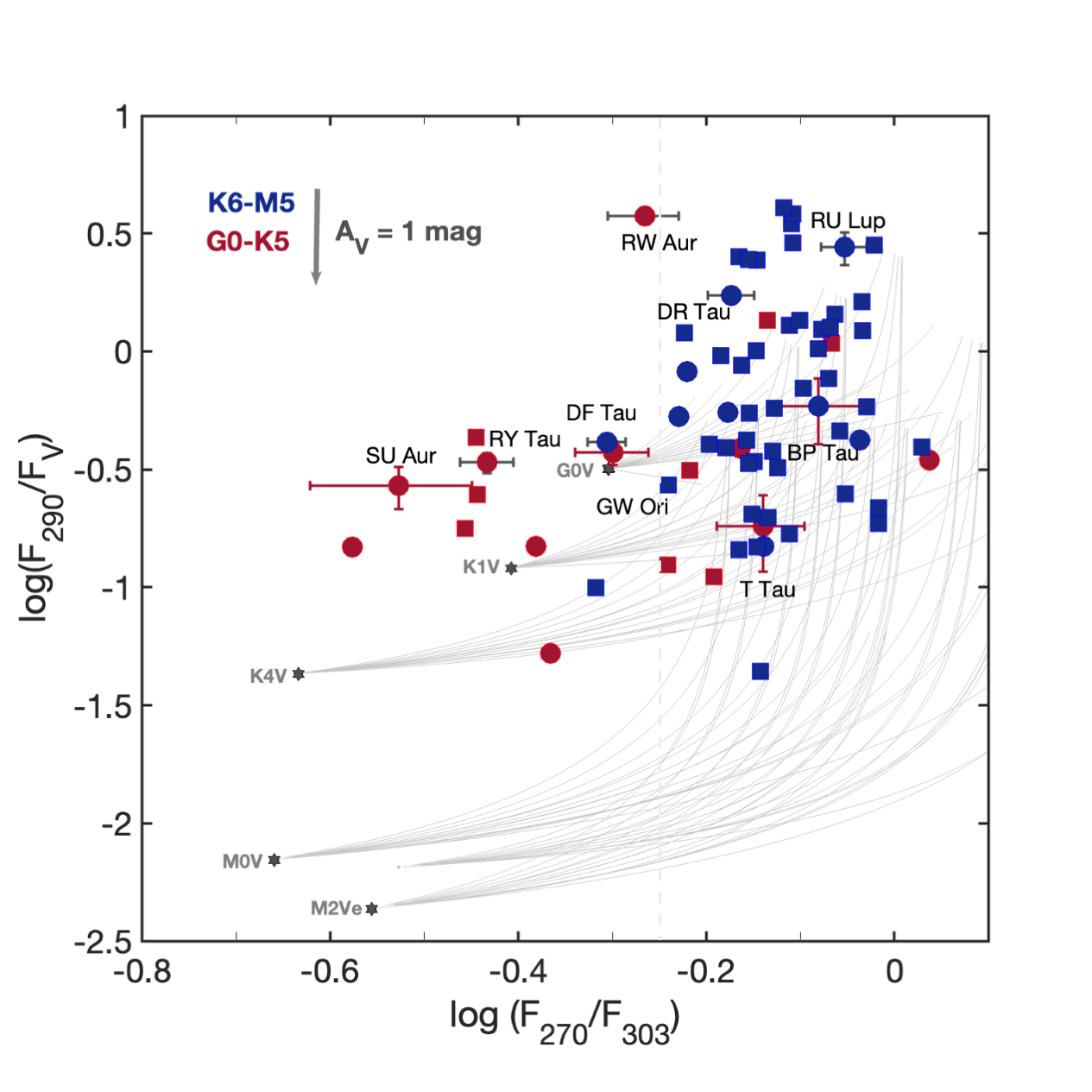}
\caption{UV colour map of accreting TTSs. Late-type stars (K6--M5) are shown in blue and warm stars (G0--K5) in red. Circles correspond to IUE observations and squares to HST/ULLYSES data. Grey curves show representative tracks from the photosphere-plus-shock model grid  adding accretion-shock continua to empirical stellar photospheres.  Labels identify stars with multiple observations. Black error bars illustrate correlated UV--$V$ variability, while red error bars represent the corresponding 1$\sigma$ uncorrelated variability. The grey arrow indicates the reddening vector for $A_V=1$ mag.}
\label{fig:models}
\end{figure}

The long temporal baseline provided by the IUE archive shows that the observed separation cannot be attributed to stellar variability. Although individual stars display substantial changes in UV and optical fluxes, the fluxes at 270 and 303\,nm remain tightly correlated, so that the $\log(F270/F303)$ colour varies only weakly over timescales ranging from days to decades. Consequently, repeated observations trace nearly parallel displacements in the colour--colour plane rather than erasing the separation between warm and late-type stars.

The distinction between the two populations becomes more evident when the observations are compared with the empirical photospheric sequence and the grid of photosphere-plus-shock models (Fig.~3). The majority of late  K--M stars lie within or above the range spanned by the adopted photosphere-plus-shock grid  and exhibit the UV excesses expected for actively accreting systems. In contrast, many warm G--K5 stars cluster close to the empirical photospheric sequence, and several are even redder than non-accreting stars of the same spectral type. These colours extend beyond the range sampled by the adopted model grid, indicating that the UV continua of warm TTSs differ systematically from those of fully convective objects.
This observational distinction forms the basis of the interpretation discussed below.

\begin{figure}
\centering
\includegraphics[width=0.50\textwidth]{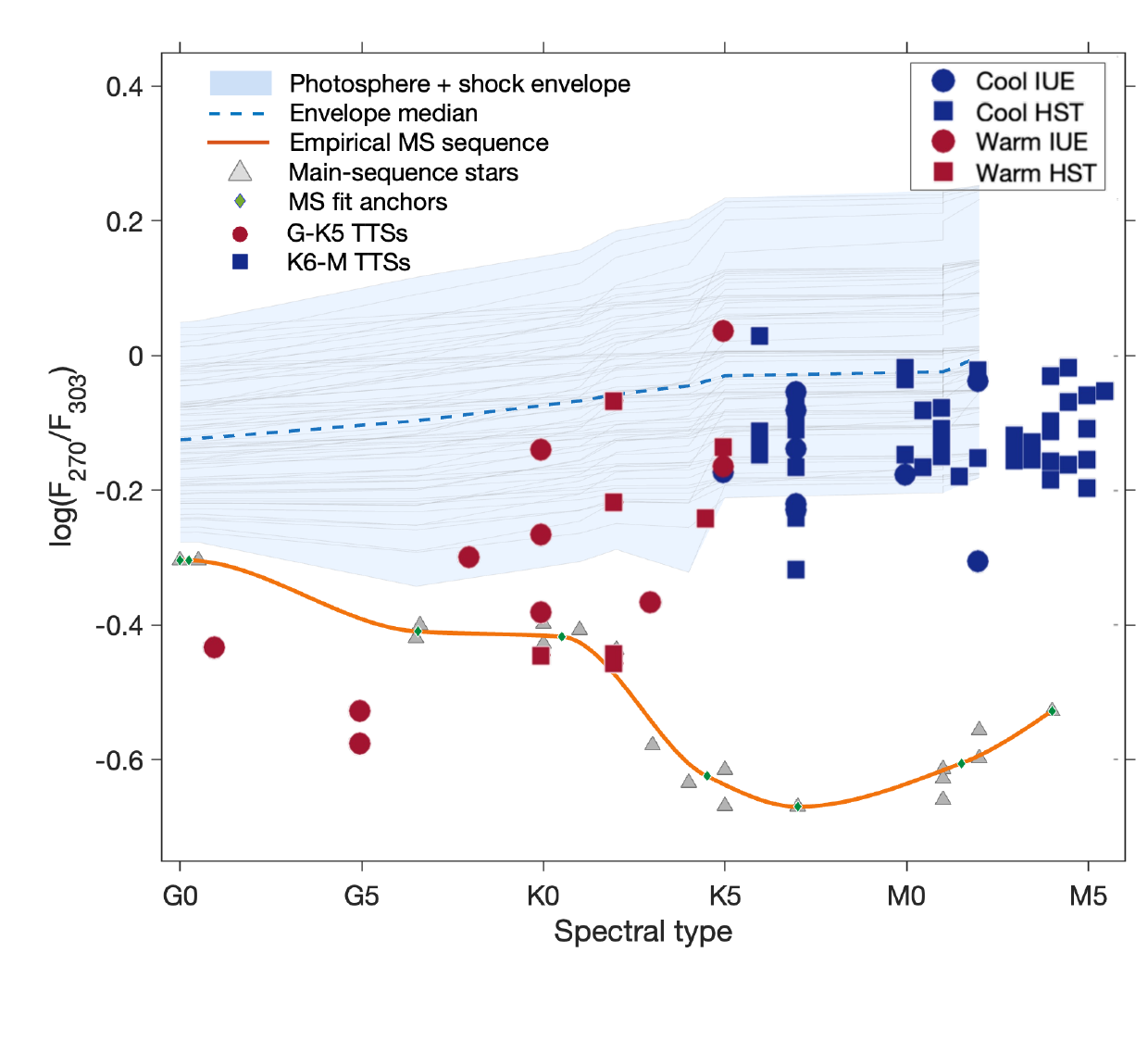}
\caption{Empirical UV photospheric sequence and photosphere-plus-shock model envelope. Warm TTSs (G0--K5; red circles), late-type TTSs (K6--M5; blue squares), and main-sequence stars (grey triangles) are shown in the $\log(F270/F303)$--spectral-type plane. The solid blue curve represents the empirical photospheric sequence derived from the main-sequence sample. The shaded region shows the range of UV colours spanned by the adopted  ensemble of photosphere-plus-shock models for veiling factors $R=0.01$--0.5. 
}
\label{fig:models}
\end{figure}

\section{Statistical analysis}

To quantify the observed separation between warm G--K5 and late K6--M stars, each object was treated as an independent measurement of the UV colour $\log(F270/F303)$. Late K--M stars exhibit mean colours 0.17 dex larger than the G--K5 population, with a bootstrap 95\% confidence interval of 0.09--0.25 dex. The difference is highly significant according to Welch's unequal-variance t-test ($p=5.5\times10^{-4}$) and the non-parametric Mann--Whitney U test ($p=6.7\times10^{-5}$). Permutation tests based on random reassignment of spectral-type labels likewise reject the null hypothesis of no association between UV colour and spectral type ($p<10^{-5}$). The result remains robust when the boundary separating warm and late-type stars is varied between spectral types K3 and K7.

To determine whether the observed behaviour is better described by a continuous trend or by a transition in UV properties, three models were considered: a single linear relation, a continuous broken-line model and a step-transition model (Fig.~4). Model comparison using the Bayesian Information Criterion (BIC) strongly favours the two transition models over a single linear trend. The broken-line model provides the best description of the observations ($\Delta {\rm BIC} \simeq 7.3$ relative to the linear model), closely followed by the step-transition model ($\Delta{\rm BIC}\simeq 6.6$). Both models locate the transition near spectral type K5--K6.

\begin{figure}
\centering
\includegraphics[width=0.50\textwidth]{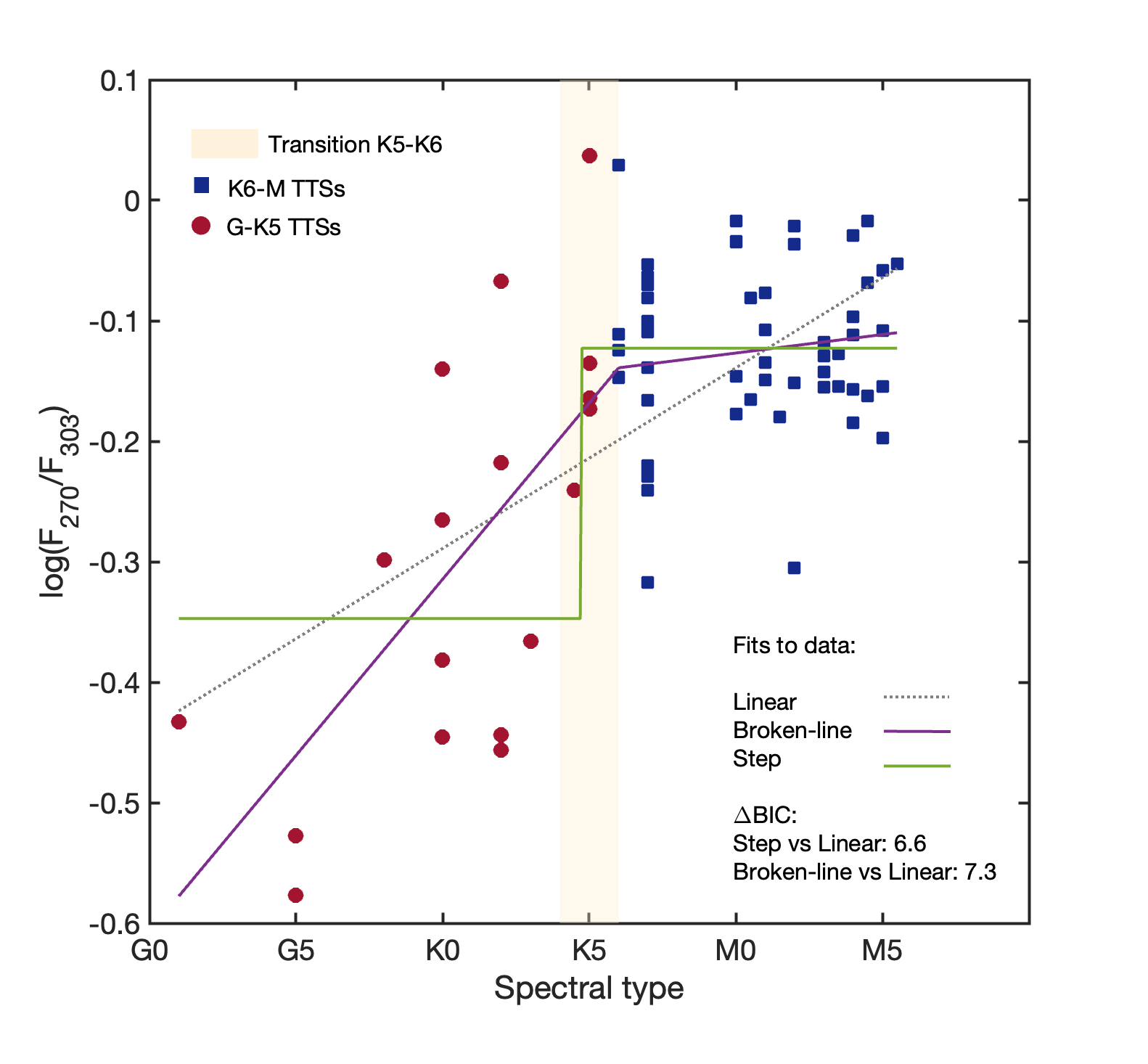}
\caption{Statistical evidence for a transition in UV continuum properties. Dependence of $\log(F270/F303)$ on spectral type for the full TTS sample. Late-type stars (K6--M5) are shown as blue squares and warm stars (G0--K5) as red circles. The curves show three statistical descriptions of the data: a single linear relation (grey dotted line), a continuous broken-line transition (purple), and a step-transition model (green). The shaded region marks the approximate spectral-type interval where pre-main-sequence stars develop radiative cores. 
}
\label{fig:models}
\end{figure}

The spectral classifications were also examined as a potential source of uncertainty in the location of the inferred transition. The ULLYSES compilation combines classifications from multiple sources based on different methodologies; we therefore compared the adopted spectral types with the recent classifications from the PENELLOPE collaboration \citep{2021A&A...650A.196M}. The two sets show good overall agreement, with the largest discrepancies concentrated around K7--M0.5. Importantly, Sz82 is the only object that changes sides of the adopted K5--K6 boundary, being classified as K5 in the ULLYSES compilation and K7 by \citet{2025ApJ...992..134P}. Repeating the analysis with Sz82 classified as K7 leaves the result essentially unchanged: the preference of the continuous broken-line model over a single linear relation changes only from $\Delta{\rm BIC}=7.35$ to 7.09, and the inferred breakpoint remains near K6.

As a more general sensitivity test, we performed 10,000 Monte Carlo realizations in which the spectral classifications of all objects were independently perturbed by up to 0.5, 1, and 2 spectral subtypes. These perturbations are intended as a sensitivity test rather than as a formal model of the spectral-type uncertainties. For perturbations of up to 0.5 subtype, the median breakpoint remains at K6, with a 95\% range of approximately K5.5--K7.9; for perturbations of up to 1 subtype, it remains at K6, with a 95\% range of K5.0--K8.2. The broken-line transition model is preferred over a single linear trend in all realizations in both cases. Even when perturbations as large as two subtypes are allowed, the median breakpoint remains close to K6 (approximately K6.3), while its 95\% range broadens to approximately K4.5--K8.6; the transition model is still preferred in 98.6\% of the realizations. Thus, increasing spectral-type uncertainty broadens the range over which the transition can be located but does not systematically shift it or remove the statistical preference for a transition.

The statistical analysis confirms that the separation identified in Figs.~2 and~3 is a population-wide property rather than the result of a few individual objects. The UV continuum does not evolve smoothly with spectral type but instead undergoes a statistically significant transition centred near K5--K6, whose precise location is broadened by uncertainties in spectral classification. The inferred transition remains consistent with the spectral range associated with the onset of radiative cores in pre-main-sequence stars.

\section{Conclusions}

We have investigated the UV continua of 68 accreting TTSs using homogeneous observations from the \textit{IUE} and the HST/ULLYSES survey. UV continuum colours reveal a clear transition near spectral type K5--K6 that is independently recovered in both datasets and is robust against extinction, long-term variability, and sample selection effects.

The statistical analysis demonstrates a transition in the UV continuum properties near spectral type K5--K6,  although spectral-classification uncertainties limit the precision with which its exact location can be determined. Comparison with photosphere-plus-shock models indicates that the largest deviations from the parameter space sampled by the adopted model grid occur preferentially among the warmer stars. Many warm G--K5 stars exhibit substantially  lower  $F270/F303$ ratios than the late-type population,and several display continua that are even redder than those of non-accreting stars of comparable spectral type. The UV continua of partially convective TTSs  therefore extend into a region of the $F270/F303$ diagnostic that is not encompassed by the adopted photosphere-plus-shock model grid.

The transition occurs close to the spectral range where pre-main-sequence stars develop radiative cores and where spectropolarimetric studies reveal major changes in magnetic topology. These results suggest that the evolution from predominantly dipolar to more complex magnetic-field configurations modifies the geometry of magnetospheric accretion, thereby altering the emergent UV continuum. Future UV spectroscopy combined with spectropolarimetric observations will help establish how changes in magnetic topology are reflected in the structure and energetics of accretion flows.

\begin{acknowledgements}
This work has received funding from the Spanish Ministry of Science, Innovation and Universities through the 2023 ''Proyectos de Generaci\'on de Conocimiento'' call, and co-funded by the European Union, under grant PID2023-147740OB-I00.
\end{acknowledgements}

\facilities{HST(STIS, COS), IUE}

\appendix

\section{Properties and UV continuum measurements of the TTSs in the sample}
\startlongtable
\begin{deluxetable*}{llcrrrrrr}
\tablecaption{Supplementary properties and ultraviolet continuum measurements of the HST/ULLYSES T Tauri star sample.\label{tab:supp_tts_hst}}
\tabletypesize{\scriptsize}
\tablewidth{0pt}
\tablehead{
\colhead{Star} &
\colhead{SpT} &
\colhead{$E(B-V)$} &
\colhead{$F_{270}$} &
\colhead{$F_{290}$} &
\colhead{$F_{303}$} &
\colhead{$F_V$} &
\colhead{$F_{270}/F_{303}$} &
\colhead{$F_{290}/F_V$}
}
\startdata
STC2dj161344m373646 & M5 & 0.194 & $4.63\times10^{-16}$ & $5.48\times10^{-16}$ & $5.94\times10^{-16}$ & $2.11\times10^{-16}$ & 0.78 & 2.59 \\
Sz104 & M5 & 0.13 & $1.16\times10^{-16}$ & $1.58\times10^{-16}$ & $1.65\times10^{-16}$ & $2.89\times10^{-16}$ & 0.70 & 0.55 \\
Sz84 & M5 & 0.19 & $1.63\times10^{-16}$ & $2.06\times10^{-16}$ & $2.56\times10^{-16}$ & $5.07\times10^{-16}$ & 0.64 & 0.41 \\
RECX9 & M4.5 & 0.01 & $2.52\times10^{-16}$ & $2.89\times10^{-16}$ & $2.62\times10^{-16}$ & $1.33\times10^{-15}$ & 0.96 & 0.22 \\
Sz10 & M4.5 & 0.258 & $2.06\times10^{-16}$ & $2.52\times10^{-16}$ & $2.99\times10^{-16}$ & $4.84\times10^{-16}$ & 0.69 & 0.52 \\
Sz69 & M4.5 & 0.1 & $1.56\times10^{-16}$ & $1.84\times10^{-16}$ & $1.83\times10^{-16}$ & $1.45\times10^{-16}$ & 0.85 & 1.26 \\
Sz103 & M4 & 0.226 & $7.60\times10^{-17}$ & $1.11\times10^{-16}$ & $1.09\times10^{-16}$ & $4.13\times10^{-16}$ & 0.70 & 0.27 \\
Sz110 & M4 & 0.29 & $8.46\times10^{-16}$ & $9.30\times10^{-16}$ & $1.06\times10^{-15}$ & $1.33\times10^{-15}$ & 0.80 & 0.70 \\
Sz76 & M4 & 0.16 & $1.89\times10^{-16}$ & $1.93\times10^{-16}$ & $2.45\times10^{-16}$ & $1.30\times10^{-15}$ & 0.77 & 0.15 \\
Sz97 & M4 & 0.19 & $4.19\times10^{-16}$ & $4.59\times10^{-16}$ & $4.48\times10^{-16}$ & $7.89\times10^{-16}$ & 0.93 & 0.58 \\
Sz99 & M4 & 0.39 & $1.90\times10^{-16}$ & $2.40\times10^{-16}$ & $2.91\times10^{-16}$ & $2.50\times10^{-16}$ & 0.65 & 0.96 \\
XXCha & M3.5 & 0.323 & $3.16\times10^{-15}$ & $3.87\times10^{-15}$ & $4.52\times10^{-15}$ & $3.00\times10^{-15}$ & 0.70 & 1.29 \\
CVSO176 & M3 & 0.1 & $2.71\times10^{-16}$ & $3.12\times10^{-16}$ & $3.64\times10^{-16}$ & $8.23\times10^{-16}$ & 0.74 & 0.38 \\
RECX6 & M3 & 0.08 & $1.10\times10^{-16}$ & $1.30\times10^{-16}$ & $1.53\times10^{-16}$ & $2.97\times10^{-15}$ & 0.72 & 0.04 \\
Sz66 & M3 & 0.11 & $1.34\times10^{-16}$ & $1.66\times10^{-16}$ & $1.91\times10^{-16}$ & $9.40\times10^{-16}$ & 0.70 & 0.18 \\
WZCha & M3 & 0.387 & $9.51\times10^{-16}$ & $1.13\times10^{-15}$ & $1.25\times10^{-15}$ & $6.00\times10^{-16}$ & 0.76 & 1.88 \\
Sz130 & M2 & 0.22 & $2.50\times10^{-16}$ & $2.88\times10^{-16}$ & $3.54\times10^{-16}$ & $1.40\times10^{-15}$ & 0.71 & 0.21 \\
Sz72 & M2 & 0.242 & $2.94\times10^{-15}$ & $3.39\times10^{-15}$ & $3.08\times10^{-15}$ & $1.95\times10^{-15}$ & 0.95 & 1.74 \\
Sz71 & M1.5 & 0.161 & $5.40\times10^{-16}$ & $6.67\times10^{-16}$ & $8.16\times10^{-16}$ & $2.35\times10^{-15}$ & 0.66 & 0.28 \\
RXJ1556m3655 & M1 & 0.323 & $6.87\times10^{-15}$ & $7.99\times10^{-15}$ & $8.81\times10^{-15}$ & $5.29\times10^{-15}$ & 0.78 & 1.51 \\
SSTC2dj161243m381503 & M1 & 0.258 & $3.30\times10^{-16}$ & $4.08\times10^{-16}$ & $4.50\times10^{-16}$ & $3.45\times10^{-15}$ & 0.73 & 0.12 \\
Sz111 & M1 & 0.32 & $6.58\times10^{-16}$ & $8.58\times10^{-16}$ & $9.27\times10^{-16}$ & $2.51\times10^{-15}$ & 0.71 & 0.34 \\
Sz45 & M0.5 & 0.226 & $2.20\times10^{-15}$ & $2.53\times10^{-15}$ & $2.65\times10^{-15}$ & $3.87\times10^{-15}$ & 0.83 & 0.66 \\
VZCha & M0.5 & 0.387 & $8.46\times10^{-15}$ & $1.06\times10^{-14}$ & $1.24\times10^{-14}$ & $9.07\times10^{-15}$ & 0.68 & 1.16 \\
CVSO104 & M0+M2 & 0.14 & $1.44\times10^{-15}$ & $1.59\times10^{-15}$ & $1.72\times10^{-15}$ & $1.28\times10^{-15}$ & 0.84 & 1.24 \\
CVSO109 & M0 & 0.15 & $4.32\times10^{-15}$ & $4.70\times10^{-15}$ & $4.67\times10^{-15}$ & $3.85\times10^{-15}$ & 0.93 & 1.22 \\
CVSO109a & M0 & 0.15 & $4.12\times10^{-15}$ & $4.50\times10^{-15}$ & $4.47\times10^{-15}$ & $2.78\times10^{-15}$ & 0.92 & 1.62 \\
CVSO109b & M0 & 0.15 & $1.93\times10^{-16}$ & $1.99\times10^{-16}$ & $2.01\times10^{-16}$ & $1.07\times10^{-15}$ & 0.96 & 0.19 \\
VWCha\_a & M0 & 0.613 & $3.18\times10^{-15}$ & $3.99\times10^{-15}$ & $4.45\times10^{-15}$ & $5.56\times10^{-15}$ & 0.71 & 0.72 \\
CVSO107 & K7 & 0.103 & $1.64\times10^{-15}$ & $1.89\times10^{-15}$ & $1.90\times10^{-15}$ & $1.62\times10^{-15}$ & 0.86 & 1.17 \\
CVSO58 & K7 & 0.039 & $6.80\times10^{-16}$ & $7.43\times10^{-16}$ & $7.99\times10^{-16}$ & $1.05\times10^{-15}$ & 0.85 & 0.71 \\
CVSO90 & K7 & 0.17 & $3.59\times10^{-15}$ & $4.38\times10^{-15}$ & $4.62\times10^{-15}$ & $1.26\times10^{-15}$ & 0.78 & 3.48 \\
SYCha & K7 & 0.161 & $8.34\times10^{-15}$ & $1.09\times10^{-14}$ & $1.40\times10^{-14}$ & $1.25\times10^{-14}$ & 0.60 & 0.87 \\
Sz129 & K7 & 0.29 & $3.96\times10^{-15}$ & $4.49\times10^{-15}$ & $4.99\times10^{-15}$ & $5.94\times10^{-15}$ & 0.79 & 0.76 \\
Sz77 & K7 & 0.2 & $1.36\times10^{-15}$ & $1.79\times10^{-15}$ & $1.99\times10^{-15}$ & $1.24\times10^{-14}$ & 0.68 & 0.14 \\
Sz98 & K7 & 0.323 & $6.59\times10^{-16}$ & $9.50\times10^{-16}$ & $1.15\times10^{-15}$ & $6.67\times10^{-15}$ & 0.57 & 0.14 \\
CVSO165 & K6 & 0.06 & $1.11\times10^{-15}$ & $1.34\times10^{-15}$ & $1.47\times10^{-15}$ & $4.18\times10^{-15}$ & 0.75 & 0.32 \\
CVSO165a & K6 & 0.11 & $4.19\times10^{-16}$ & $5.36\times10^{-16}$ & $5.88\times10^{-16}$ & $3.62\times10^{-15}$ & 0.71 & 0.15 \\
CVSO165b & K6 & 0.4 & $6.84\times10^{-16}$ & $8.09\times10^{-16}$ & $8.84\times10^{-16}$ & $6.28\times10^{-16}$ & 0.77 & 1.29 \\
Sz75 & K6 & 0.226 & $1.79\times10^{-14}$ & $2.26\times10^{-14}$ & $2.51\times10^{-14}$ & $3.52\times10^{-14}$ & 0.71 & 0.64 \\
V505Ori & K6 & 0.3 & $2.80\times10^{-16}$ & $2.96\times10^{-16}$ & $2.62\times10^{-16}$ & $7.52\times10^{-16}$ & 1.07 & 0.39 \\
Sz82 & K5 & 0.29 & $6.86\times10^{-15}$ & $8.39\times10^{-15}$ & $9.37\times10^{-15}$ & $1.11\times10^{-14}$ & 0.73 & 0.76 \\
TXOri & K4.5 & 0.4 & $6.08\times10^{-16}$ & $9.82\times10^{-16}$ & $1.06\times10^{-15}$ & $7.92\times10^{-15}$ & 0.57 & 0.12 \\
CHX18n & K2 & 0.258 & $1.12\times10^{-15}$ & $1.97\times10^{-15}$ & $2.28\times10^{-15}$ & $1.65\times10^{-14}$ & 0.49 & 0.12 \\
RXJ0438p1546 & K2 & 0.097 & $5.33\times10^{-15}$ & $1.40\times10^{-14}$ & $1.48\times10^{-14}$ & $6.87\times10^{-14}$ & 0.36 & 0.20 \\
RXJ1842m3532 & K2 & 0.129 & $1.31\times10^{-14}$ & $1.46\times10^{-14}$ & $1.53\times10^{-14}$ & $1.73\times10^{-14}$ & 0.86 & 0.84 \\
RXJ1852m3700 & K2 & 0.323 & $1.84\times10^{-15}$ & $3.00\times10^{-15}$ & $3.04\times10^{-15}$ & $1.83\times10^{-14}$ & 0.61 & 0.16 \\
SZCha & K2 & 0.42 & $4.89\times10^{-16}$ & $1.13\times10^{-15}$ & $1.40\times10^{-15}$ & $1.48\times10^{-14}$ & 0.35 & 0.08 \\
Sz19 & K0 & 0.484 & $5.21\times10^{-15}$ & $1.14\times10^{-14}$ & $1.45\times10^{-14}$ & $6.89\times10^{-14}$ & 0.36 & 0.17 \\
\enddata
\tablecomments{Fluxes are continuum flux densities measured in 5\,nm-wide windows centred at 270, 290, and 303\,nm and in the optical $V$ band. Flux densities are given in units of erg\,s$^{-1}$\,cm$^{-2}$\,\AA$^{-1}$.}
\end{deluxetable*}

\startlongtable
\begin{deluxetable}{llcrrc}
\tablecaption{Supplementary ultraviolet continuum colours of the IUE T Tauri star sample.\label{tab:supp_tts_iue}}
\tabletypesize{\scriptsize}
\tablewidth{0pt}
\tablehead{
\colhead{Star} &
\colhead{SpT} &
\colhead{$E(B-V)$} &
\colhead{$F_{270}/F_{303}$} &
\colhead{$F_{290}/F_V$} &
\colhead{$N_{\rm obs}$}
}
\startdata
DETau & M2 & 0.29 & 0.92 & 0.24 & 6 \\
DFTau & M2 & 0.19 & 0.50 & 0.41 & 10 \\
DL Tau & M0 & 0.42 & 0.66 & 0.24 & 1 \\
GMAur & K7 & 0.19 & 0.73 & 0.10 & 2 \\
AA Tau & K7 & 0.613 & 0.59 & 0.16 & 1 \\
DGTau & K7 & 0.48 & 0.60 & 0.32 & 7 \\
BPTau & K7 & 0.35 & 0.83 & 0.29 & 38 \\
RULup & K7 & 0.22 & 0.89 & 1.78 & 22 \\
DSTau & K5 & 0.13 & 0.69 & 0.30 & 3 \\
DK Tau & K5 & 0.39 & 1.09 & 0.16 & 1 \\
DRTau & K5 & 0.452 & 0.67 & 0.70 & 35 \\
V410Tau & K3 & 0.12 & 0.43 & 0.04 & 2 \\
ABDor & K0 & 0 & 0.42 & 0.15 & 8 \\
TTau & K0 & 0.29 & 0.72 & 0.10 & 10 \\
RWAur & K0 & 1 & 0.54 & 0.51 & 23 \\
GWOri & G8 & 0.45 & 0.50 & 0.15 & 9 \\
HD283572 & G5 & 0.097 & 0.40 & 0.15 & 5 \\
SUAur & G5 & 0.29 & 0.30 & 0.15 & 20 \\
RYTau & G1 & 0.71 & 0.37 & 0.08 & 60 \\
\enddata
\tablecomments{The IUE colour indices were derived from repeated ultraviolet and optical observations when available. $N_{\rm obs}$ gives the number of IUE observations used for each object.}
\end{deluxetable}

\clearpage
\section{Properties and UV continuum measurements of the main sequence stars comparison sample}

\begin{deluxetable*}{lccrrrrrr}[h!]
\tablecaption{IUE main-sequence comparison sample.\label{tab:ms_iue}}
\tablehead{
\colhead{Star} &
\colhead{SpT} &
\colhead{$d$} &
\colhead{$F_{270}$} &
\colhead{$F_{290}$} &
\colhead{$F_{303}$} &
\colhead{$F_V$} &
\colhead{$F_{270}/F_{303}$} &
\colhead{$F_{290}/F_V$} \\
\colhead{} &
\colhead{} &
\colhead{(pc)} &
\colhead{} &
\colhead{} &
\colhead{} &
\colhead{} &
\colhead{} &
\colhead{}
}
\startdata
HD119850 & M4e & 16.74 & $2.18\times10^{-15}$ & $9.38\times10^{-15}$ & $7.33\times10^{-15}$ & $4.27\times10^{-13}$ & 0.297 & 0.0219 \\
HD95735 & M2b & 2.55 & $5.62\times10^{-15}$ & $2.07\times10^{-14}$ & $2.22\times10^{-14}$ & $3.07\times10^{-13}$ & 0.253 & 0.0676 \\
HD217987 & M2b & 3.29 & $8.91\times10^{-15}$ & $2.43\times10^{-14}$ & $3.2\times10^{-14}$ & $6.06\times10^{-13}$ & 0.278 & 0.0401 \\
HD233153 & M1 & 12.28 & $1.14\times10^{-15}$ & $4.17\times10^{-15}$ & $4.69\times10^{-15}$ & $4.12\times10^{-13}$ & 0.244 & 0.0101 \\
HD202560 & M1 & 3.97 & $1.97\times10^{-14}$ & $5.84\times10^{-14}$ & $9.01\times10^{-14}$ & $1.31\times10^{-12}$ & 0.219 & 0.0444 \\
HD103932 & K5 & 10.17 & $3.4\times10^{-14}$ & $1.17\times10^{-13}$ & $1.4\times10^{-13}$ & $6.77\times10^{-12}$ & 0.243 & 0.0173 \\
HD209100 & K4 & 3.64 & $3.83\times10^{-13}$ & $1.39\times10^{-12}$ & $1.65\times10^{-12}$ & $4.27\times10^{-12}$ & 0.232 & 0.3245 \\
HD22049 & K2 & 3.22 & $3.6\times10^{-12}$ & $9.99\times10^{-12}$ & $9.84\times10^{-12}$ & $1.31\times10^{-11}$ & 0.366 & 0.7600 \\
HD17925 & K1 & 10.36 & $4.67\times10^{-13}$ & $1.23\times10^{-12}$ & $1.19\times10^{-12}$ & $1.09\times10^{-11}$ & 0.391 & 0.1123 \\
HD115617 & G6.5 & 8.53 & $3.64\times10^{-12}$ & $8.41\times10^{-12}$ & $9.55\times10^{-12}$ & $3.93\times10^{-11}$ & 0.381 & 0.2137 \\
HD72905 & G0.5 & 14.44 & $3.19\times10^{-12}$ & $5.77\times10^{-12}$ & $6.43\times10^{-12}$ & $3.79\times10^{-11}$ & 0.496 & 0.1523 \\
HD109358 & G0 & 8.47 & $1.32\times10^{-11}$ & $2.35\times10^{-11}$ & $2.66\times10^{-11}$ & $5.28\times10^{-11}$ & 0.497 & 0.4457 \\
\enddata
\tablecomments{Fluxes are continuum flux densities measured in the 5\,nm windows centred at 270, 290, and 303\,nm, and in the $V$ band, in units of erg\,s$^{-1}$\,cm$^{-2}$\,\AA$^{-1}$.}
\end{deluxetable*}

\begin{deluxetable*}{lccrrrr}
\tablecaption{HST main-sequence comparison sample.\label{tab:ms_hst}}
\tablehead{
\colhead{Star} &
\colhead{SpT} &
\colhead{$d$} &
\colhead{$F_{270}$} &
\colhead{$F_{290}$} &
\colhead{$F_{303}$} &
\colhead{$F_{270}/F_{303}$} \\
\colhead{} &
\colhead{} &
\colhead{(pc)} &
\colhead{} &
\colhead{} &
\colhead{} &
\colhead{}
}
\startdata
GJ5825 & M1 & 3.97 & $2.42\times10^{-14}$ & $7.09\times10^{-14}$ & $1.03\times10^{-13}$ & 0.236 \\
HR8086 & K7 & 3.5 & $4.22\times10^{-14}$ & $1.37\times10^{-13}$ & $1.97\times10^{-13}$ & 0.214 \\
HD201091 & K5 & 3.5 & $1.37\times10^{-13}$ & $4.7\times10^{-13}$ & $6.4\times10^{-13}$ & 0.214 \\
HD122064 & K3 & 10.07 & $9.05\times10^{-14}$ & $3\times10^{-13}$ & $3.43\times10^{-13}$ & 0.264 \\
HD010780 & K0 & 10.04 & $1.46\times10^{-12}$ & $3.66\times10^{-12}$ & $3.91\times10^{-12}$ & 0.374 \\
HD185144 & K0 & 5.76 & $4\times10^{-12}$ & $9.31\times10^{-12}$ & $9.99\times10^{-12}$ & 0.400 \\
HD115617 & G6.5 & 8.53 & $5.02\times10^{-12}$ & $1.11\times10^{-11}$ & $1.26\times10^{-11}$ & 0.398 \\
\enddata
\tablecomments{Fluxes are continuum flux densities measured in the 5\,nm windows centred at 270, 290, and 303\,nm, in units of erg\,s$^{-1}$\,cm$^{-2}$\,\AA$^{-1}$.}
\end{deluxetable*}

\bibliography{bibliography.bib}{}

\begin{thebibliography}{}
\expandafter\ifx\csname natexlab\endcsname\relax\def\natexlab#1{#1}\fi
\providecommand{\url}[1]{\href{#1}{#1}}
\providecommand{\dodoi}[1]{doi:~\href{http://doi.org/#1}{\nolinkurl{#1}}}
\providecommand{\doeprint}[1]{\href{http://ascl.net/#1}{\nolinkurl{http://ascl.net/#1}}}
\providecommand{\doarXiv}[1]{\href{https://arxiv.org/abs/#1}{\nolinkurl{https://arxiv.org/abs/#1}}}

\bibitem[{F. {Allard} {et~al.}(2012){Allard}, {Homeier}, \&
  {Freytag}}]{2012RSPTA.370.2765A}
{Allard}, F., {Homeier}, D., \& {Freytag}, B. 2012, \bibinfo{title}{{Models of
  very-low-mass stars, brown dwarfs and exoplanets},} Philosophical
  Transactions of the Royal Society of London Series A, 370, 2765,
  \dodoi{10.1098/rsta.2011.0269}

\bibitem[{J. {Bouvier} {et~al.}(2007){Bouvier}, {Alencar}, {Harries},
  {Johns-Krull}, \& {Romanova}}]{2007prpl.conf..479B}
{Bouvier}, J., {Alencar}, S.~H.~P., {Harries}, T.~J., {Johns-Krull}, C.~M., \&
  {Romanova}, M.~M. 2007, \bibinfo{title}{{Magnetospheric Accretion in
  Classical T Tauri Stars},} in Protostars and Planets V, ed. B.~{Reipurth},
  D.~{Jewitt}, \& K.~{Keil}, 479, \dodoi{10.48550/arXiv.astro-ph/0603498}

\bibitem[{A. {Dodin}(2018){Dodin}}]{2018MNRAS.475.4367D}
{Dodin}, A. 2018, \bibinfo{title}{{The structure and spectrum of the accretion
  shock in the atmospheres of young stars},} \mnras, 475, 4367,
  \dodoi{10.1093/mnras/sty038}

\bibitem[{J.-F. {Donati} {\&} J.~D. {Landstreet}(2009){Donati} \&
  {Landstreet}}]{2009ARA&A..47..333D}
{Donati}, J.-F., \& {Landstreet}, J.~D. 2009, \bibinfo{title}{{Magnetic Fields
  of Nondegenerate Stars},} \araa, 47, 333,
  \dodoi{10.1146/annurev-astro-082708-101833}

\bibitem[{J.-F. {Donati} {et~al.}(2024){Donati}, {Finociety}, {Cristofari},
  {Alencar}, {Moutou}, {Delfosse}, {Fouqu{\'e}}, {Arnold}, {Baruteau},
  {K{\'o}sp{\'a}l}, {M{\'e}nard}, {Carmona}, {Grankin}, {Takami}, {Artigau},
  {Doyon}, {H{\'e}brard}, \& {the SPIRou science team}}]{2024MNRAS.530..264D}
{Donati}, J.-F., {Finociety}, B., {Cristofari}, P.~I., {et~al.} 2024,
  \bibinfo{title}{{The classical T Tauri star CI Tau observed with SPIRou:
  magnetospheric accretion and planetary formation},} \mnras, 530, 264,
  \dodoi{10.1093/mnras/stae675}

\bibitem[{A.~I. {G{\'o}mez de Castro} {\&} M. {Fern{\'a}ndez}(1996){G{\'o}mez
  de Castro} \& {Fern{\'a}ndez}}]{1996MNRAS.283...55G}
{G{\'o}mez de Castro}, A.~I., \& {Fern{\'a}ndez}, M. 1996,
  \bibinfo{title}{{Ultraviolet spectroscopy of the hotspot in the classical T
  Tauri star DI Cep: observational indications of magnetically channelled
  accretion.},} \mnras, 283, 55, \dodoi{10.1093/mnras/283.1.55}

\bibitem[{A.~I. {G\'omez de Castro} {et~al.}(1997){G\'omez de Castro},
  {Franqueira}, \& {Wamsteker}}]{1997iueu.book.....G}
{G\'omez de Castro}, A.~I., {Franqueira}, M., \& {Wamsteker}, W. 1997,
  {International Ultraviolet Explorer--Uniform Low Dispersion Archive : T Tauri
  stars}, Vol.~8

\bibitem[{A.~I. {G\'omez de Castro} {\&} S.~A. {Lamzin}(1999){G\'omez de
  Castro} \& {Lamzin}}]{1999MNRAS.304L..41G}
{G\'omez de Castro}, A.~I., \& {Lamzin}, S.~A. 1999, \bibinfo{title}{{Accretion
  shocks in T Tauri stars: diagnosis via semiforbidden ultraviolet line
  ratios},} \mnras, 304, L41, \dodoi{10.1046/j.1365-8711.1999.02545.x}

\bibitem[{S.~G. {Gregory} {et~al.}(2016){Gregory}, {Adams}, \&
  {Davies}}]{2016MNRAS.457.3836G}
{Gregory}, S.~G., {Adams}, F.~C., \& {Davies}, C.~L. 2016, \bibinfo{title}{{The
  influence of radiative core growth on coronal X-ray emission from
  pre-main-sequence stars},} \mnras, 457, 3836, \dodoi{10.1093/mnras/stw259}

\bibitem[{S.~G. {Gregory} {et~al.}(2012){Gregory}, {Donati}, {Morin},
  {Hussain}, {Mayne}, {Hillenbrand}, \& {Jardine}}]{2012ApJ...755...97G}
{Gregory}, S.~G., {Donati}, J.-F., {Morin}, J., {et~al.} 2012,
  \bibinfo{title}{{Can We Predict the Global Magnetic Topology of a
  Pre-main-sequence Star from Its Position in the Hertzsprung-Russell
  Diagram?},} \apj, 755, 97, \dodoi{10.1088/0004-637X/755/2/97}

\bibitem[{E. {Gullbring} {et~al.}(2000){Gullbring}, {Calvet}, {Muzerolle}, \&
  {Hartmann}}]{2000ApJ...544..927G}
{Gullbring}, E., {Calvet}, N., {Muzerolle}, J., \& {Hartmann}, L. 2000,
  \bibinfo{title}{{The Structure and Emission of the Accretion Shock in T Tauri
  Stars. II. The Ultraviolet-Continuum Emission},} \apj, 544, 927,
  \dodoi{10.1086/317253}

\bibitem[{A. {Guseva} {et~al.}(2026){Guseva}, {Manchon}, {Petitdemange}, \&
  {Pin{\c{c}}on}}]{2026A&A...707A..45G}
{Guseva}, A., {Manchon}, L., {Petitdemange}, L., \& {Pin{\c{c}}on}, C. 2026,
  \bibinfo{title}{{Radial differential rotation leading to dipole collapse in
  pre-main-sequence stars},} \aap, 707, A45,
  \dodoi{10.1051/0004-6361/202557616}

\bibitem[{L. {Hartmann} {et~al.}(2016){Hartmann}, {Herczeg}, \&
  {Calvet}}]{2016ARA&A..54..135H}
{Hartmann}, L., {Herczeg}, G., \& {Calvet}, N. 2016, \bibinfo{title}{{Accretion
  onto Pre-Main-Sequence Stars},} \araa, 54, 135,
  \dodoi{10.1146/annurev-astro-081915-023347}

\bibitem[{L. {Hartmann} {et~al.}(1994){Hartmann}, {Hewett}, \&
  {Calvet}}]{1994ApJ...426..669H}
{Hartmann}, L., {Hewett}, R., \& {Calvet}, N. 1994,
  \bibinfo{title}{{Magnetospheric Accretion Models for T Tauri Stars. I. Balmer
  Line Profiles without Rotation},} \apj, 426, 669, \dodoi{10.1086/174104}

\bibitem[{L. {Ingleby} {et~al.}(2013){Ingleby}, {Calvet}, {Herczeg}, {Blaty},
  {Walter}, {Ardila}, {Alexander}, {Edwards}, {Espaillat}, {Gregory},
  {Hillenbrand}, \& {Brown}}]{2013ApJ...767..112I}
{Ingleby}, L., {Calvet}, N., {Herczeg}, G., {et~al.} 2013,
  \bibinfo{title}{{Accretion Rates for T Tauri Stars Using Nearly Simultaneous
  Ultraviolet and Optical Spectra},} \apj, 767, 112,
  \dodoi{10.1088/0004-637X/767/2/112}

\bibitem[{C.~M. {Johns-Krull}(2007){Johns-Krull}}]{2007ApJ...664..975J}
{Johns-Krull}, C.~M. 2007, \bibinfo{title}{{The Magnetic Fields of Classical T
  Tauri Stars},} \apj, 664, 975, \dodoi{10.1086/519017}

\bibitem[{A. {Koenigl}(1991){Koenigl}}]{1991ApJ...370L..39K}
{Koenigl}, A. 1991, \bibinfo{title}{{Disk Accretion onto Magnetic T Tauri
  Stars},} \apjl, 370, L39, \dodoi{10.1086/185972}

\bibitem[{M. {Long} {et~al.}(2007){Long}, {Romanova}, \&
  {Lovelace}}]{2007MNRAS.374..436L}
{Long}, M., {Romanova}, M.~M., \& {Lovelace}, R.~V.~E. 2007,
  \bibinfo{title}{{Accretion to stars with non-dipole magnetic fields},}
  \mnras, 374, 436, \dodoi{10.1111/j.1365-2966.2006.11192.x}

\bibitem[{C.~F. {Manara} {et~al.}(2021){Manara}, {Frasca}, {Venuti}, {Siwak},
  {Herczeg}, {Calvet}, {Hernandez}, {Tychoniec}, {Gangi}, {Alcal{\'a}},
  {Boffin}, {Nisini}, {Robberto}, {Briceno}, {Campbell-White},
  {Sicilia-Aguilar}, {McGinnis}, {Fedele}, {K{\'o}sp{\'a}l}, {{\'A}brah{\'a}m},
  {Alonso-Santiago}, {Antoniucci}, {Arulanantham}, {Bacciotti}, {Banzatti},
  {Beccari}, {Benisty}, {Biazzo}, {Bouvier}, {Cabrit}, {Caratti o Garatti},
  {Coffey}, {Covino}, {Dougados}, {Eisl{\"o}ffel}, {Ercolano}, {Espaillat},
  {Erkal}, {Facchini}, {Fang}, {Fiorellino}, {Fischer}, {France}, {Gameiro},
  {Garcia Lopez}, {Giannini}, {Ginski}, {Grankin}, {G{\"u}nther}, {Hartmann},
  {Hillenbrand}, {Hussain}, {James}, {Koutoulaki}, {Lodato}, {Mauc{\'o}},
  {Mendigut{\'\i}a}, {Mentel}, {Miotello}, {Oudmaijer}, {Rigliaco}, {Rosotti},
  {Sanchis}, {Schneider}, {Spina}, {Stelzer}, {Testi}, {Thanathibodee}, {Vink},
  {Walter}, {Williams}, \& {Zsidi}}]{2021A&A...650A.196M}
{Manara}, C.~F., {Frasca}, A., {Venuti}, L., {et~al.} 2021,
  \bibinfo{title}{{PENELLOPE: The ESO data legacy program to complement the
  Hubble UV Legacy Library of Young Stars (ULLYSES). I. Survey presentation and
  accretion properties of Orion OB1 and {\ensuremath{\sigma}}-Orionis},} \aap,
  650, A196, \dodoi{10.1051/0004-6361/202140639}

\bibitem[{S. {Peacock} {et~al.}(2022{\natexlab{a}}){Peacock}, {Barman},
  {Schneider}, {Leung}, {Schwieterman}, {Shkolnik}, \&
  {Loyd}}]{2022ApJ...933..235P}
{Peacock}, S., {Barman}, T.~S., {Schneider}, A.~C., {et~al.}
  2022{\natexlab{a}}, \bibinfo{title}{{Accurate Modeling of
  Ly{\ensuremath{\alpha}} Profiles and Their Impact on Photolysis of
  Terrestrial Planet Atmospheres},} \apj, 933, 235,
  \dodoi{10.3847/1538-4357/ac77f2}

\bibitem[{C.~V. {Pittman} {et~al.}(2025){Pittman}, {Espaillat}, {Robinson},
  {Thanathibodee}, {Lopez}, {Calvet}, {Zhu}, {Walter}, {Wendeborn}, {Manara},
  {Campbell-White}, {Claes}, {Fang}, {Frasca}, {Gameiro}, {Gangi},
  {Hern{\'a}ndez}, {K{\'o}sp{\'a}l}, {Mauc{\'o}}, {Muzerolle}, {Siwak},
  {Tychoniec}, \& {Venuti}}]{2025ApJ...992..134P}
{Pittman}, C.~V., {Espaillat}, C.~C., {Robinson}, C.~E., {et~al.} 2025,
  \bibinfo{title}{{The ODYSSEUS Survey. Characterizing Magnetospheric
  Geometries and Hotspot Structures in T Tauri Stars},} \apj, 992, 134,
  \dodoi{10.3847/1538-4357/adef35}

\bibitem[{J. {Roman-Duval} {et~al.}(2025){Roman-Duval}, {Fischer}, {Fullerton},
  {Taylor}, {Plesha}, {Proffitt}, {Monroe}, {Fischer}, {Aloisi}, {Bouret},
  {Britt}, {Calvet}, {Carlberg}, {Crowther}, {De Rosa}, {Dixon}, {Espaillat},
  {Evans}, {Fox}, {France}, {Garcia}, {Fleming}, {Frazer}, {G{\'o}mez de
  Castro}, {Herczeg}, {Hernandez}, {Hirschauer}, {James}, {Johns-Krull},
  {Leitherer}, {Lockwood}, {Najita}, {Oey}, {Oliveira}, {Pauly}, {Reid},
  {Riedel}, {Rodriguez}, {Sahnow}, {Sankrit}, {Sembach}, {Shaw}, {Smith},
  {Sohn}, {Som}, {{\'U}beda}, \& {Welty}}]{2025ApJ...985..109R}
{Roman-Duval}, J., {Fischer}, W.~J., {Fullerton}, A.~W., {et~al.} 2025,
  \bibinfo{title}{{The UV Legacy Library of Young Stars as Essential Standards
  (ULLYSES) Large Director's Discretionary Program with Hubble. I. Goals,
  Design, and Initial Results},} \apj, 985, 109,
  \dodoi{10.3847/1538-4357/adc45b}

\end{thebibliography}
\bibliographystyle{aasjournalv7.1}

\end{document}